\documentclass[aps,prd,preprint,superscriptaddress,nofootinbib]{revtex4}
\usepackage{amsmath,amssymb,enumerate}

\newcommand{\bignone}{}
\newcommand{\emdash}{---}
\newcommand{\mathd}{\mathrm{d}}
\newcommand{\tmem}[1]{{\em #1\/}}
\newcommand{\tmop}[1]{\ensuremath{\operatorname{#1}}}
\newcommand{\tmstrong}[1]{\textbf{#1}}
\newenvironment{enumeratenumeric}{\begin{enumerate}[1.] }{\end{enumerate}}

\begin{document}

\title{(Not) Summing over Kaluza-Kleins}

\author{Johannes Hirn}
\email{johannes.hirn@yale.edu}
\affiliation{Department of Physics, Yale University, New Haven, CT
  06520}
\author{Ver\'onica Sanz}
\email{vsanz@bu.edu}
\affiliation{Department of Physics, Boston University, Boston, MA 02215}

\begin{abstract}
  Models in extra-dimensions have unique features. Many of their surprising
  properties simply result from the underlying 5D structure. This  structure
  shows up as ``Sum Rules'' involving the whole tower of Kaluza-Kleins. In
  this paper, we present a holographic {\tmem{shortcut}} and derive these
  results without solving the eigenvalue problem: we express 4D physical
  quantities directly in terms of the 5D metric. In warped space, one can go
  further and isolate the effect of the new physics sector. This method can be
  used for any 5D model, and we apply it here to the case of holographic QCD
  and technicolor.
\end{abstract}

\maketitle

\section{Introduction}

We are interested in physical predictions of 5D models. These are often
derived using the Kaluza-Klein (KK) decomposition. The heavy KKs can be
integrated out, and one obtains the effective theory valid at low energies.
Physical quantities in this effective theory involve contributions from{\tmem{
all}} modes. Such relations expressing 4D observables as a sum over KKs are
called {\tmem{Sum Rules}} (SRs). Computing these SRs is in general difficult: it
involves solving the eigenvalue problem for a --large-- number of states.

Here we present a method to bypass this problem. One can obtain each 4D
physical quantity from a single geometrical computation in 5D, illustrating an
application of holography. In practice, this is possible because KKs are
excitations of a single 5D field, and thus obey properties of orthogonality
and completeness. Using these properties, one transforms the SRs into
geometrical factors.

This method has clear advantages besides computational economy: {\tmem{1)}}
computations can be carried out for all metrics, not just flat space or
$\tmop{AdS}$, {\tmem{2)}} relations among 4D observables (which depend on the 5D
structure of the model) show up with no need of specifying the metric, and
finally {\tmem{3)}} one can obtain approximate expressions for classes of
metrics. Since physically successful geometries seem to require something
beyond flat space or even pure AdS, point {\tmem{1)}} above turns out to be
particularly useful.

In the present article, we describe a wide range of new applications, in order
to demonstrate the generality and the power of the method. We focus on the 5D
case as opposed to the deconstructed one, in order to work with differential
equations rather than matrices. The results would go through for moose models.
To our knowledge, the first SRs were derived in deconstructed models in
connection with the Weinberg SRs of QCD {\cite{hep-ph/0304182}} and in 5D
models in connection with the high-energy behavior of longitudinal $W W$
scattering {\cite{hep-ph/0305237}} {\footnote{See also
{\cite{hep-ph/0111016,hep-ph/0201164}}.}}. Other applications to
deconstruction can be found in
{\cite{hep-ph/0401032,hep-ph/0403112,hep-ph/0406077,hep-ph/0408067}}, and to
extra-dimensional models in
{\cite{hep-ph/0507049,hep-th/0507073,hep-ph/0606086,hep-ph/0612070}}.

In Section 2, we demonstrate the technique in a simple case: the boundary
conditions (BCs) select from the 5D field only massive spin-1 fields. From
this, the reader should be able to apply the method to other situations. In
Section 3, we collect and comment some results that can be obtained. We then
turn to electroweak applications by coupling the model to $\tmop{SU} \left( 2
\right) \times \mathrm{U} \left( 1 \right)$ gauge fields in Section 4. We
illustrate the role played by the warp factor, and use our method to construct
an expansion for the masses. In Section 5, we apply our methods to the rho and
$T$ parameters. We conclude in Section 6.

\section{What are sum rules good for?}\label{what}

To give a flavor of the way things work, we derive SRs in a simple setting.
Physical applications will follow in the next Sections. The method can be
summarized as follows,
\begin{eqnarray}
  \text{4D quantities} & = & \sum_{\tmop{KKs}} \bignone \text{KK properties} 
  \nonumber\\
  & = &  \text{Geometrical factor} .  \label{geom}
\end{eqnarray}
The extra dimension considered here is an interval. The two ends of the space
are located at $z = l_0$ (the {\tmem{UV brane}}) and $z = l_1$ (the {\tmem{IR
brane}}). We define the metric as follows
\begin{eqnarray}
  \mathd s^2 & = & w \left( z \right)^2  \left( \eta_{\mu \nu} \mathd x^{\mu}
  \mathd x^{\nu} - \mathd z^2 \right),  \label{2.2}
\end{eqnarray}
where $w (z)$ is the{\tmem{ warp factor}}. The {\tmem{geometrical factors}} of
(\ref{geom}) are multiple integrals of $w (z)$ {\tmem{only}}, and can thus be
computed directly once the metric is specified.

For example $w (z) = 1, l_0 / z$ represent flat space and AdS
{\cite{hep-ph/9905221}} respectively. {\tmem{Gap metrics}} are metrics whose
$w$ grows in the UV as least as fast as AdS {\cite{hep-ph/0606086}}. For
applications to electroweak symmetry breaking (EWSB), this growth induces a
gap between the light sector ($\gamma, Z, W$) and the KKs. We will see the
importance of gap metrics in Section \ref{virtues}.

For this Section and the next, we consider fields with $\left( - \right)$ UV
BCs. Only starting from Section 4 will we consider other UV BCs.

\subsection{An example: KK masses}\label{exampleKK}

Let us give a first example of a SR. To compute some physical observables we
need to obtain the effect of KK exchange. This exchange is suppressed by the
KK masses, and we want to compute the following sum
\begin{eqnarray}
  \bignone \Sigma & \equiv & \sum_{n = 1}^{\infty}  \frac{1}{M_n^2}
  \hspace{1em} = \hspace{1em} ?  \label{SR1}
\end{eqnarray}
Such a sum can be calculated for very simple metrics only. In flat space,
considering only the first KK in the sum (\ref{SR1}) would provide an
approximation good at the 20\% level. For AdS, the sum is still dominated by
the first resonance, although the error is this time of 30\%.

Apart from exceptional cases, one cannot obtain an analytic expression for the
spectrum. Moreover, in such cases, a large number of KKs can give sizeable
contributions to the sum. Should we solve numerically for a large number of
eigenmodes? The answer is no, since we can obtain the geometrical factor in
the right-hand side of the SR (\ref{SR1}) in 3 steps :

{\tmem{Step 1.)}} Use the normalization of the wave-functions to write
\begin{eqnarray}
  \sum_n \frac{1}{M_n^2} & = & \frac{2}{g_5^2} \sum_n \frac{\int w \varphi_n^2
  \bignone}{M_n^2} .  \label{exampleSR}
\end{eqnarray}
Note the shorthand notation of Appendix \ref{notations}, very useful
throughout the paper. The aim is then to get rid of the $1 / M_n^2$ factor in order
to apply the completeness relation. \

{\tmem{Step 2.)}} This is achieved by integrating twice the equation of
motion (EOM), which we write as
\begin{eqnarray}
  \frac{1}{M_n^2} \partial \left( w \partial \varphi_n \right) & = & - w
  \varphi_n .  \label{EOM}
\end{eqnarray}
To integrate this, attention must be paid to the BCs. Sticking for the moment
to the simplest case of $\left( -, + \right)$ BCs, integration of the EOMs
(\ref{EOM}) leads to
\begin{eqnarray}
  \frac{\varphi_n \left( z \right)}{M_n^2} & = & \int^z \frac{1}{w}  \int_{z'}
  w \varphi_n .  \label{phi/M}
\end{eqnarray}
{\tmem{Step 3.)}} We can then plug the expression (\ref{phi/M}) into the sum
(\ref{exampleSR}), and use the completeness relation {\footnote{This
distribution is to be applied to a function satisfying the same BCs as
$\varphi$.}}
\begin{eqnarray}
  \frac{2}{g_5^2} w \left( z \right)  \sum_n \varphi_n \left( z \right)
  \varphi_n \left( z' \right) & = & \delta \left( z - z' \right) . 
  \label{kpl}
\end{eqnarray}
For $\left( + \right)$ \ IR BCs one obtains for the SR (\ref{SR1}) \
\begin{eqnarray}
  \Sigma_+ \hspace{1em} \equiv \hspace{1em} \sum_n \frac{1}{M_{+, n}^2} & = &
  \int w \int^z \frac{1}{w} . \hspace{2em} \text{{\tmstrong{SR1}}} 
  \label{resSR1}
\end{eqnarray}
For fields with $(-)$ IR BCs, the result would be
\begin{eqnarray}
  \Sigma_- \hspace{1em} = \hspace{1em} \sum_n \frac{1}{M_{-, n}^2} & = & \int
  w \alpha_- \int^z \frac{1}{w}, \hspace{2em} \text{{\tmstrong{SR1'}}}  
  \label{resSR1p}
\end{eqnarray}
where $\alpha_-$ is itself a function of $w$
\begin{eqnarray}
  \alpha_- \left( z \right) & = & \frac{\int_z  \frac{1}{w} \bignone}{\int 
  \frac{1}{w} \bignone} .  \label{ainf}
\end{eqnarray}
$\Sigma_+$ and $\Sigma_-$ are derived assuming $(-)$ UV BCs. We can read off
the SR for $(+)$ UV BCs by just swapping $l_0$ and $l_1$. A field satisfying
$(+, -)$ BCs follows the SR
\begin{eqnarray}
  \Sigma_+^s & = & \Sigma_+ (l_0 \leftrightarrow l_1) .
  \hspace{2em} \text{{\tmstrong{SR1''}}}  \label{swapping}
\end{eqnarray}
Finally, the SR for a $(+, +)$ field is divergent, signaling the presence of a
massless mode.

This method can also be applied to mixed BCs, see Appendix \ref{App-mixed}
for details. Keep also in mind that the expressions
(\ref{resSR1}-\ref{swapping}) are valid for gauge bosons. For other spins, the
derivation can be performed along the same lines.

\subsection{KK exchange}\label{KKexch}

Consider now a more physical quantity: the contribution of KK exchange to some
process. Assume the KKs are coupled to some other fields (fermion bilinears
with profile $f$) via the overlap integral
\begin{eqnarray}
  g_n & = & \int \frac{\mathd z}{g_5^2} w \left( z \right) f \left( z \right)
  \varphi_n \left( z \right) \bignone, 
\end{eqnarray}
for the example of spin-1 KKs with $\left( + \right)$ IR BCs. The contribution
of KK exchange to 4-fermion interactions can then be expressed as
\begin{eqnarray}
  \sum_n \frac{g_n^2}{M_n^2} & = & \frac{1}{2 g_5^2}  \int wf \int^z
  \frac{1}{w}  \int_{z'} wf .  \label{gnM2}
\end{eqnarray}
This is done again by using first the implicit expression for $\varphi / M^2$
(\ref{phi/M}) in terms of a double integral and then applying completeness
relation (\ref{kpl}).

If the fermions are localized at a particular point in the fifth dimension,
$z_{\ast}$, one obtains
\begin{eqnarray}
  \sum_n \frac{\varphi_n \left( z_{\ast} \right)^2}{M_n^2} & = &
  \frac{g_5^2}{2}  \int^{z_{\ast}} \frac{1}{w} . \hspace{2em}
  \text{{\tmstrong{SR2}}}  \label{exKKex}
\end{eqnarray}

Note also that there may be situations in which the infinite sums
{\tmstrong{SR1}} and {\tmstrong{SR1'}} (\ref{SR1}) do not converge. For
example, in the setting of {\cite{hep-ph/0602229}}, masses follow Regge
trajectories $M_n^2 \sim n$ \ and {\tmstrong{SR1}} diverges. On the other hand
the physical quantity (\ref{gnM2}) is finite since $f \left( + \infty \right)
= 0$, and so is the SR (\ref{exKKex}).

\section{The case of Holographic QCD }

Let us describe some SRs and their meaning in a simple example, Holographic
QCD. The bulk gauge symmetry is the QCD chiral symmetry, $\tmop{SU} \left( N_f
\right)_L \times \tmop{SU} \left( N_f \right)_R$ , with a common 5D coupling
$g_5$. \

Chiral symmetry is broken spontaneously, leading to massless pions at low
energies. The spontaneous breaking has a clear interpretation in an AdS
background. Breaking of chiral symmetry by the vev of an operator (condensate)
of dimension $d$ is done by introducing a bifundamental, $X (z)$, whose vev is
the order parameter for chiral symmetry breaking
{\cite{hep-ph/0501128,hep-ph/0501218}}
\begin{eqnarray}
  \langle X \rangle & = & \sigma z^d .  \label{bgfield}
\end{eqnarray}
The value of $d$ is set by the $X$ bulk mass: $m^2 / l_0^2 = d (d - 4)$
{\footnote{For a review on the AdS/CFT correspondence, see for example
{\cite{hep-th/9905111}} and references therein.}}.

At the quadratic level, $X$ only couples to the axial combination of the
$\tmop{LR}$ bulk fields. This effect can be recast as an {\tmem{effective}}
metric for the axial field {\cite{hep-ph/0612239}}, as well as a modification
of the IR BC. The effect of these modified BCs is explained in Appendix
\ref{App-mixed}. Therefore, whereas the vector field only{\tmem{ sees}} the
AdS background,
\begin{eqnarray}
  w_V (z) & = & \frac{l_0}{z}, 
\end{eqnarray}
the axial sees a different metric
\begin{eqnarray}
  w_A (z) & = & \frac{l_0}{z} f (z) .  \label{axialw}
\end{eqnarray}
The particular form of $f (z)$ depends on the dimension of the operator. We
gave details on how to compute its form in {\cite{hep-ph/0612239}}.

Other background fields can also act on vector and axial fields without
breaking chiral symmetry. In that case, $w_V \left( z \right) \neq l_0 / z$.
For the present paper, one only needs to take into account that in general the
{\tmem{effective}} metrics of vector and axial are different
\begin{eqnarray}
  w_V & \neq & w_A . 
\end{eqnarray}
The key point here is that the coupling of $A$ or $V$ to background fields can
be recast as an effective metric and different BCs. And that is all you need
to extract SRs.

Within this setup, one can obtain many SRs involving resonance couplings
($g_n, f_n, \alpha_n$) and the low energy constants ($L_i$'s) of the chiral Lagrangian {\footnote{See {\cite{hep-ph/0507049}} for definitions.
One can also extract SRs involving the $\mathcal{O}(p^6)$ Lagrangian
 along the lines of Section \ref{KKexch}. }}. Let us enumerate
some of them,
\begin{eqnarray}
  \sum_{n = 1}^{\infty} f_{V_n} g_{V_n} M_{V_n}^2 & = f_{\pi}^2 & =
  \frac{1}{g_5^2  \int \bignone \frac{1}{w_A}} , \hspace{2em} \text{{\tmstrong{SR3}}}  
  \label{fpi}
\end{eqnarray}
\begin{eqnarray}
  \sum_{n = 1}^{\infty} f_{V_n} g_{V_n} & = & 2 L_9 = \frac{1}{2 g_5^2}  \int
  w_V  \left( 1 - \alpha_A^2 \right) , \hspace{2em} \text{{\tmstrong{SR4}}}  \label{fVgV}\\
  \sum_{n = 1}^{\infty} g_{V_n}^2 & = & 8 L_1 = \frac{1}{4 g_5^2}  \int w_V
  \left( 1 - \alpha_A^2 \right)^2  , \hspace{2em} \text{{\tmstrong{SR5}}}  \label{L1SR}\\
  &  &  \nonumber\\
  \sum_{n = 1}^{\infty} f_{A_n}  \left( f_{A_n} + 2 \sqrt{2} \alpha_{A_n}
  \right) & = & 4 \left( L_9 + L_{10} \right) = \frac{1}{g_5^2}  \int
  \alpha_A^2  \left( w_A - w_V  \right)  , \hspace{2em}  \text{{\tmstrong{SR6}}} 
  \label{L910SR}
\end{eqnarray}
What is the physical importance of these SRs? All them ensure the
{\tmem{high-energy softness}} of some physical process. This behavior has been
experimentally tested and/or is expected from theoretical grounds. In
resonance models, these relations have to be imposed by hand. In a 5D model it
comes automatically from the non-locality in the 5th dimension and the 5D
gauge invariance.

{\tmstrong{SR3}} and {\tmstrong{SR4}} ensure high-energy softness of the
vector form factor. {\tmstrong{SR5}} protects the elastic Goldstone boson
scattering from violating the Froissart bound. {\tmstrong{SR6}} is essential
for the soft high energy behavior of the axial form factor. Summarizing,
physical amplitudes are determined by the following SRs
\begin{eqnarray*}
  & \tmop{High} \tmop{Energy} \tmop{Softness} & 
\end{eqnarray*}
\begin{eqnarray*}
  \tmop{Unitarity}, \tmop{GB} \tmop{scattering} & \Longrightarrow &
  \text{{\tmstrong{SR5}}} ,\\
  \tmop{Vector} \tmop{Form} \tmop{Factor} & \Longrightarrow &
  \text{{\tmstrong{SR3, SR4}}} ,\text{}\\
  \tmop{Axial} \tmop{Form} \tmop{Factor} & \Longrightarrow &
  \text{{\tmstrong{SR6}}} ,
\end{eqnarray*}
Two comments are in order:
\begin{enumeratenumeric}
  \item These SRs and consequently, the correct behavior of some amplitudes at
  high energy, are satisfied {\tmem{whatever the form of the effective metrics
  $w_A$ and $w_V$}}.
  
  \item It is impossible to verify these relations by {\tmem{summing over a
  finite number of KKs}}.
\end{enumeratenumeric}
This gives a very clear example of how (not) summing over KKs is useful to get
physical quantities.

Another interesting SR is the one for the $S$-parameter,
\begin{eqnarray}
  S & = & 4 \pi \sum_{n = 1}^{\infty} \left( f_{V_n}^2 - f_{A_n}^2 \right) 
  \label{fV2-fA2}\\
  & = & \frac{4 \pi}{g_5^2}  \int w_V - w_A \alpha_A^2  . \hspace{2em}  \text{{\tmstrong{SR7}}} \nonumber
\end{eqnarray}
The detailed derivation of this SR is done in {\cite{hep-ph/0612239}}. We used
the same simple techniques of Section \ref{exampleKK} to obtain
{\tmstrong{SR7}}.

Why is it important that the SRs are satisfied {\tmem{whatever the form of
the metric}}? Effective metrics leading to interesting phenomenology for
electroweak symmetry breaking are not simply flat or AdS. Gauge fields couple
to background fields as in (\ref{bgfield}), representing electroweak symmetry
breaking or dilaton couplings. Using our method {\cite{hep-ph/0612239}}, one
can rewrite these fields as effective metrics (\ref{axialw}). The resulting
metric is no longer AdS, but a more complicated function.
{\tmstrong{SR7}} is a compact and exact relation between the $S$ parameter and
the effective metric, whatever its form.

Another advantage of {\tmstrong{SR7}} is that computing partial sums over the
first $f_{V, A}$'s in (\ref{fV2-fA2}) would give an incorrect answer. For
non-flat or AdS metrics, the sum in (\ref{fV2-fA2}) is not saturated by the
first resonance  . One should then sum over a number of KKs (still below the
IR cutoff) to obtain a good approximation for $S$. {\tmstrong{SR7}} bypasses
this problem by giving at once the exact answer.

\section{The many virtues of gap metrics}\label{virtues}

In this section we describe the importance of working with gap metrics to turn
a QCD-like model into a description of EWSB.

\subsection{The meaning of the UV brane: the holographic
recipe}\label{UVbrane}

We have described how the bulk gauge symmetry is broken near the IR. To solve
the second-order EOMs, one must specify a second set of BCs: those on the UV
brane.

To select these BCs, one evaluates the action on the solution, by plugging in
the EOMs. For a Yang-Mills field, there remains a UV boundary term
\begin{eqnarray}
  \mathcal{S} & \propto & \left. \frac{1}{g_5^2} V_{\mu} \left( x, z \right)
  \partial_z V_{\mu} \left( x, z \right) \right|_{z = l_0} +\mathcal{O} \left(
  V^3 \right) .  \label{VdV}
\end{eqnarray}
In applications to QCD, one fixed the UV boundary value of the field $V_{\mu}
(x, l_0) \equiv v_{\mu} (x)$. $v_{\mu} \left( x \right)$ enjoys the properties
of a source for the currents of the 4D global symmetry $\mathcal{G} \supset
\text{$\tmop{SU} \left( N_f \right)_L \times \tmop{SU} \left( N_f
\right)_R$}$. In that case, (\ref{VdV}) is rewritten as
\begin{eqnarray}
  &  & \left. \frac{1}{g_5^2} v_{\mu} \left( x \right) \partial_z V_{\mu}
  \left( x, z \right) \right|_{z = l_0}, 
\end{eqnarray}
we can identify the symmetry currents of $\mathcal{G}$ {\footnote{This
expression can be decomposed as a sum over KK modes: see \cite{hep-ph/0507049}
for a discussion. }}
\begin{eqnarray}
  J_{\mu} & = & \frac{1}{g_5^2} \partial_z V_{\mu} |_{l_0} +\mathcal{O} \left(
  V^2 \right) . 
\end{eqnarray}
Since one sets the sources equal to zero in order to extract Green's
functions, the fields have Dirichlet $\left( - \right)$ UV BCs in that case.

For applications to technicolor, one couples $\tmop{SU} \left( 2 \right)
\times \mathrm{U} \left( 1 \right)$ gauge fields to the conserved currents of
$\mathcal{G}$. In the holographic version, one promotes an $\tmop{SU} \left( 2
\right) \times \mathrm{U} \left( 1 \right)$ subgroup of $\mathcal{G}$ to be
dynamical: the UV boundary values should be {\tmem{independent}} degrees of
freedom. When decomposed in terms of KK modes, the boundary term (\ref{VdV})
would then mix the new {\tmem{dynamical}} source with the resonances,
analogously to $\rho - \gamma$ mixing in QCD. To diagonalize the action, the
UV BCs should be changed from Dirichlet to Neumann
\begin{eqnarray}
  \partial_z V_{\mu} (x, z) |_{z = l_0}  & = & 0, 
\end{eqnarray}
for the $\tmop{SU} \left( 2 \right) \times \mathrm{U} \left( 1 \right)$
subgroup. The boundary term (\ref{VdV}) then vanishes.

The procedure described above, of switching BCs to go from Holographic QCD to
Holographic Technicolor, is valid irrespective of the metric. However, the
change of BCs generally modifies the eigenmodes: only for gap metrics do the KK modes
for the technicolor BCs keep a memory of the spectrum with QCD BCs.

To describe the change from $(-)$ to $(+)$ UV BCs one can use
{\tmstrong{SR1'}} and {\tmstrong{SR1''}}. We define the gap ($G$) as the ratio
of the sums $\Sigma$ for these two cases
\begin{eqnarray}
  G & = & \frac{\Sigma_+^s}{\Sigma_-} .  \label{G}
\end{eqnarray}
Gap metrics satisfy
\begin{eqnarray}
  G \gg & 1 & , 
\end{eqnarray}
and the value of $G$ is UV-dependent. For example, in AdS,
\begin{eqnarray}
  G & = & \log \left( \frac{l_1^2}{l_0^2} \right) . 
\end{eqnarray}
$G \gg 1$ signals the appearance of a new sector in the model: the
{\tmem{ultra-light}} sector (UL). Whatever the metric, the sources acquire a
mass as they become dynamical. The difference is that, in gap metrics, this
mass is fed mainly to the UL sector, not the KKs. We will explain this point
in detail in the next two subsections.

\subsection{The gap as KK repellent}\label{repelled}

A gap metric is a metric which grows as $1 / z$ or faster near the UV. In that
case $\int w \bignone$ contains a parametrically large factor, which diverges
for $l_0 \longrightarrow 0$. This is going to impact the normalization of
states, as well as the geometrical factors appearing in SRs, since these are
integrals of $w$. For practical applications, we are interested in the
limiting case of metrics that are approximately AdS near the UV, and our
$G$-expansion will apply to that case for simplicity {\footnote{The
$G$-expansion can be performed for other gap metrics, but the order at which
corrections appear would be different.}}.

Following Section \ref{UVbrane}, we now study the eigenmodes for $(+)$ BCs on
the UV brane. \ Consider removing the UV cut-off, i.e. $l_0 \longrightarrow
0$. If a mode $\varphi$ has to be normalizable
\begin{eqnarray}
  \int_0 w \varphi^2 & < & + \infty, 
\end{eqnarray}
then it must satisfy
\begin{eqnarray}
  \varphi \left( 0 \right) & = & 0 . 
\end{eqnarray}
We recognize a $\left( - \right)$ BC, even though we started out by asking
that $\varphi$ satisfy a $\left( + \right)$ BC.

Next, we turn to what happens when $l_0 > 0$. The wave-function obtains
corrections suppressed in the expansion in $1 / G$. Indeed, one can show that
the value of the field on the UV brane is suppressed
\begin{eqnarray}
  \varphi \left( l_0 \right) \hspace{1em} = \hspace{1em} \mathcal{O} \left(
  G^{- 1} \right) & \ll & \varphi \left( z \sim l_1 \right) \hspace{1em} =
  \hspace{1em} \mathcal{O} \left( G^0 \right) . 
\end{eqnarray}
The mass of such modes is given in units of $l_1$, the only remaining scale in
the boundary value problem
\begin{eqnarray}
  M_{\tmop{KK}}^2 & \propto & \frac{1}{l_1^2} . 
\end{eqnarray}

A bonus of this effect is the following: since these resonances do not
distinguish Dirichlet from Neumann BCs in the UV, they are insensitive to the
breaking happening there. In the case of Holographic Technicolor, where the UV
breaking is of $\tmop{SU} \left( 2 \right)_R \times \mathrm{U} \left( 1
\right)_{B - L}$ to $\mathrm{U} \left( 1 \right)_Y$, the spectrum of the KK
modes is approximately isospin symmetric, i.e. the masses of the fist two
isospin components (in the $W$ tower) are the same as the masses of the KKs of
the third component (the excitations of the $Z$), up to $\mathcal{O} \left( 1
/ G \right)$ corrections.

Summarizing, in gap metrics the KK sector satisfies approximately $(-)$ UV
BCs.

\subsection{Ultra-light modes in the generic case}\label{ULgen}

Let's go back to the limit $l_0 \longrightarrow 0$ and study the
non-normalizable modes. Massless modes satisfy the following EOM
\begin{eqnarray}
  \partial \left( w \partial \alpha) \right. & = & 0 .  \label{EOMalpha}
\end{eqnarray}
Solutions to this equation are
\begin{eqnarray}
  \partial \alpha & \propto \hspace{1em} 0 \hspace{1em} \tmop{or} \hspace{1em}
  & \frac{1}{w} .  \label{dalph}
\end{eqnarray}
For $\left( +, + \right)$ BCs, the solution is a flat mode, independently of
the metric. For other BCs, the second solution in (\ref{dalph}) is the right
one. For example, for $(-, -)$ BCs, $\alpha$ is given by Equation
(\ref{ainf}). In Appendix \ref{App-mixed} we give more details on the solution
 $\alpha$ for mixed IR BCs.

As Eq.(\ref{dalph}) shows, $\alpha$ will satisfy $\left( + \right)$ UV BC in
any metric where $w \left( z \right) \longrightarrow \infty$ in the UV. In gap
metrics, the mode is then flat near the UV {\footnote{In flat space, it would
be a linearly decreasing mode. In deconstructed spaces, this is the breathing
mode with an enhanced contribution on the site corresponding to the UV brane
(which has a smaller gauge coupling). We thank Andy Cohen, Shelly Glashow,
Thomas Gr\'egoire and Claudio Rebbi for a discussion of this point.}}. This is
important: when $l_0 > 0$, the norm of its kinetic term will still be large
(compared to its mass term). To get to a canonical normalization, the
wave-function has to be rescaled to
\begin{eqnarray}
  \varphi_{\tmop{UL}} & = & \mathcal{O} \left( G^{- 1 / 2} \right) . 
\end{eqnarray}
Because of this, such modes remain light compared to the KKs,
\begin{eqnarray}
  \frac{M_{\tmop{UL}}^2}{M_{\tmop{KK}}^2} & = & \mathcal{O}(G^{- 1})
  \hspace{1em} \ll \hspace{1em} 1 . 
\end{eqnarray}
This justifies our notation UL for ultra-light modes, which are kept apart
from the rest of the KK tower.

\subsection{The particular case of EWSB}

In models of EWSB, the ultra-light modes are identified with the $W$ and $Z$
(and $\gamma$). Starting with a bulk gauge group $\tmop{SU} (2)_L \times
\tmop{SU} (2)_R \times U (1)_{B - L}$, one breaks it to the electroweak by BCs
{\tmem{\`a la Higgsless}}
{\cite{hep-ph/0305237,hep-ph/0308036,hep-ph/0308038,hep-ph/0401160,hep-ph/0409126}}
taking into account the effect of bulk breaking, as described in
{\cite{hep-ph/0512240,hep-ph/0606086}}.

Rather than $L$ and $R$ fields, we work with $V$ and $A$ combinations, which
diagonalize the IR BCs
\begin{eqnarray}
  \left. A \right|_{l_1} & = & 0,  \label{IRpm1}\\
  \left. \partial V \right|_{l_1} & = & 0,  \label{IRpm2}\\
  \left. \partial B \right|_{l_1} & = & 0 . 
\end{eqnarray}
At the other end, the UV BCs are
\begin{eqnarray}
  V^{\pm} + A^{\pm} |_{l_0} & = & 0,  \label{UVpm1}\\
  \partial \left( V^{\pm} - A^{\pm} \right) |_{l_0} & = & 0,  \label{UVpm2}
\end{eqnarray}
for the first two isospin components. For the third isospin component, one
imposes
\begin{eqnarray}
  V^0 + A^0 - B |_{l_0} & = & 0, \\
  \partial \left( V^0 - A^0 \right) |_{l_0} & = & 0, \\
  \partial \left( V^0 + A^0 + \frac{\tilde{g}_5^2}{g_5^2} B \right) |_{l_0} &
  = & 0 . 
\end{eqnarray}
We will thus expand a generic wave-function in terms of its $A$ and $V$
components (and a $B$ component for the neutral modes). Let us define the
eigen-functions satisfying the appropriate BCs as two/three-component vectors
for the charged/neutral vector bosons
\begin{eqnarray}
  \hspace*{\fill} \left| \Psi_{X^{\pm}} \right\rangle & = & \left| A_{X^\pm},
  V_{X^\pm} \right\rangle,  \label{comp}\\
  \hspace*{\fill} \left| \Psi_{X^0} \right\rangle & = & \left| A_{X^0}, V_{X^0}, B_{X^0}
  \right\rangle .  \label{comp2}
\end{eqnarray}
Then the norm will be given by, \
\begin{eqnarray}
  \left\langle \Psi \left| \Psi \right\rangle \right. & \equiv & 2 \int 
  \bignone \left( \frac{w_A}{g_5^2} A^2 + \frac{w_V}{g_5^2} V^2 + \frac{w_B}{2
  \widetilde{g_5}^2} B^2 \right) .  \label{norm}
\end{eqnarray}

\subsubsection{the KK sector}

We saw in Section \ref{repelled} that, at order $G^0$, the KK sector is given
by solutions with $\left( - \right)$ UV BCs. At this order, one can thus
classify the KKs into modes that are mainly vector or mainly axial. Indeed,
for the charged sector, we have a subset of modes with a large vector
component and a small axial one
\begin{eqnarray}
  \left. | \Psi_{V^\pm_n} \right\rangle & = & \left. | 0, \varphi_{+, n}
  \right\rangle +\mathcal{O} \left( G^{- 1} \right), 
\end{eqnarray}
and vice-versa
\begin{eqnarray}
  \left. | \Psi_{A^\pm_n} \right\rangle & = & \left. | \varphi_{-, n}, 0
  \right\rangle +\mathcal{O} \left( G^{- 1} \right) . 
\end{eqnarray}
In both cases, the leading-order component is the $\varphi_{+, n}$ or
$\varphi_{-, n}$ obtained with the $\left( - \right)$ UV BC. Note that both
components are only of order $G^{- 1}$ in $l_0$ and the two components are
comparable and satisfy the UV BC: $V + A |_{l_0} = 0$.

Since the spectrum of heavy modes goes through from the case with $\left( -
\right)$ UV BCs, we can deduce the following approximate sum rule
\begin{eqnarray}
  \Sigma_W^{\tmop{KK}} \hspace{1em} \equiv \hspace{1em} \sum_{X = V, A ; n =
  1}^{\infty} \frac{1}{M_{X_n^{\pm}}^2} & = & \Sigma_+ + \Sigma_- +\mathcal{O}
  \left( G^{- 1} \right),  \label{sumKKW}
\end{eqnarray}
where $\Sigma_{+, -}$ are given in (\ref{resSR1}-\ref{resSR1p}).

The same method can be applied for the neutral KKs, where the modes split into
three towers
\begin{eqnarray}
  \left. | \Psi_{V^0_n} \right\rangle & = & \left. | 0, \varphi_{+, n}, 0
  \right\rangle +\mathcal{O} \left( G^{- 1} \right), \\
  \left. | \Psi_{A^0_n} \right\rangle & = & \left. | \varphi_{-, n}, 0, 0
  \right\rangle +\mathcal{O} \left( G^{- 1} \right), \\
  \left. | \Psi_{B^0_n} \right\rangle & = & \left. | 0, 0, \varphi_{+, n}
  \right\rangle +\mathcal{O} \left( G^{- 1} \right) . 
\end{eqnarray}
The sum of all the inverse of the masses is given by
\begin{eqnarray}
  \Sigma_{Z, \gamma}^{\tmop{KK}} \hspace{1em} \equiv \hspace{1em} \sum_{X = V,
  A, B ; n = 1}^{\infty} \frac{1}{M^2_{X^0_n}} & = & 2 \Sigma_+ + \Sigma_-
  +\mathcal{O} \left( G^{- 1} \right),  \label{sumKKZ}
\end{eqnarray}
where we have assumed for simplicity $w_B = w_V$.

\subsubsection{The ultra-light modes and their masses}

Here we illustrate the use of sum rules to extract the mass of the $W$ in an
expansion in $1 / G$. The advantage of the method is again that it will work
whatever the metric, without having to solve numerically the boundary value
problem.

Out of the ultra-light modes described in the general case in Section
\ref{ULgen}, we can construct combinations that satisfy the UV BCs for the
EWSB case. This yields the wave-functions of the $W, Z$ and $\gamma$. We get,
at order $\mathcal{O} \left( G^{- 1 / 2} \right)$
\begin{eqnarray}
  \left| W \right\rangle & \simeq & \sqrt{\frac{g_5^2}{2 Gl_0}}  \left|
  \alpha_-, - 1 \right\rangle,  \label{Wz}\\
  \left| Z \right\rangle & \simeq & \sqrt{\frac{g_5^2}{2 Gl_0}} 
  \sqrt{\frac{g_5^2 + 2 \widetilde{g_5}^2}{g_5^2 + \widetilde{g_5}^2}}  \left|
  \alpha_-, - \frac{g_5^2 }{g_5^2 + 2 \widetilde{g_5}^2}, \frac{2
  \widetilde{g_5}^2 }{g_5^2 + 2 \widetilde{g_5}^2} \right\rangle, 
  \label{Zz}\\
  \left| \gamma \right\rangle & \simeq & \sqrt{\frac{g_5^2}{2 Gl_0}}  \left|
  0, 1, 1 \right\rangle,  \label{gammaz}
\end{eqnarray}
where $\alpha_-$ is given by (\ref{ainf}).

We now want to extract the masses of these modes for a completely generic
metric. We know that the UL modes are much lighter than the other KKs. Using the same techniques as before, we can derive a sum rule
to express the sum including the UL mode
\begin{eqnarray}
  \Sigma_W & \equiv & \frac{1}{M_W^2} + \Sigma_W^{\tmop{KK}}, 
\end{eqnarray}
in terms of a geometrical factor. This sum will be dominated by the UL mode.
Similarly to {\tmstrong{SR1}}, one can derive this SR by taking into account
the difference in BCs (\ref{IRpm1}-\ref{UVpm2}). The result is
\begin{eqnarray}
  \Sigma_W & = & \int^{} w_A \int_z \frac{1}{w_A} + \int w_V \int
  \frac{1}{w_A} + \int w_V  \int^z \frac{1}{w_V} .  \label{SW}
\end{eqnarray}
In gap metrics, $\Sigma_W$ diverges as $l_0 \longrightarrow 0$ due to the
presence of the UL sector.

Using Equation (\ref{sumKKW}), one obtains that $\Sigma_W$ is also equal to $1
/ M_W^2$ at order $G^{- 1}$,
\begin{eqnarray}
  M_W^2 & = & \frac{1}{\Sigma_W} +\mathcal{O} \left( G^{- 2} \right) . 
  \label{MWleading}
\end{eqnarray}
We can go one step further in the approximation, since we know the
contribution from the other (heavy) KK modes at order $G^0$ (\ref{sumKKW}). We
can thus write
\begin{eqnarray}
  M_W^2 & = & \frac{1}{\Sigma_W - \Sigma_+ - \Sigma_-} +\mathcal{O} \left(
  G^{- 3} \right) . 
\end{eqnarray}

\subsection{Holographically turning QCD into Technicolor}

In the above, we have looked into the procedure of allowing the UV boundary
fields to be dynamical. We found an important difference between flat space
and gap metrics. In flat space, the sources would couple to {\tmem{all}} the
resonances. Therefore, when the boundary gauge fields are turned on, the whole
spectrum is changed. This can be contrasted with the case of gap metrics:
\begin{itemize}
  \item In gap metrics, the KKs are repelled from the UV brane. KKs with $(+)$
  UV BCs approximately also satisfy $(-)$ UV BCs. In gap metrics, their
  wave-functions and masses are changed only by $1 / G$ corrections
  
  \item Because of this, the KKs become approximately {\tmem{blind}} to the
  UV, and separate into modes that are essentially $V$ or essentially $A$,
  with a smaller admixture of the other. For the neutral component, there are
  also $B$ modes.
  
  \item Since isospin breaking is located on the UV brane, the KKs are
  approximately blind to it. $V$ \& $A$ resonances are thus isospin symmetric
  at leading order.
  
  \item The modes that were non-normalizable in the limit $G \longrightarrow
  \infty$ are now normalizable, but ultra-light. At leading order in $G$, the
  wave-function of the axial UL mode is proportional to the would-be
  Goldstones. These UL modes feel the UV BCs, and should be combined together in order to
  yield the physical $W$ and $Z$.
\end{itemize}
\section{Rho and $T$ parameters}\label{rho-T}

The models we consider here belong to the general class of models with
``universal corrections'', as defined in
{\cite{Barbieri:2004qk,hep-ph/0604111}}. This class encompasses the case of
oblique corrections, but is more general {\footnote{For a good illustration in
the related case of a deconstructed extra-dimension, see
{\cite{hep-ph/0406077}}.}}. In addition to the oblique corrections to $W$
and $Z$ interactions, such models also exhibit non-oblique corrections
involving the \ KK modes only.

A good example is the following: non-oblique (but universal) corrections
entail making a distinction between the rho parameter {\footnote{$\rho_{\ast}
\left( 0 \right)$ is the ratio of neutrino interactions via neutral currents
with respect to those occurring via charged currents.}} and the $T$
parameter {\footnote{$T$ only compares $W$ and $Z$ interactions.}}. In the
low-energy limit, the exchange of KK resonances generate other local
four-fermion terms in addition to the ones due to the exchange of $W$ and $Z$.
The KK-modes contribute to $\rho_{\ast} \left( 0 \right)$, but {\tmem{not}} to
$T$. Thus $\rho_{\ast} \left( 0 \right) \neq 1 + \alpha T$.

In this section we obtain a SR showing that custodial symmetry in the bulk
ensures
\begin{eqnarray}
  \rho_{\ast} (0)_{} & = & 1, 
\end{eqnarray}
for{\tmem{ any }}metric. The whole tower contributes to the SR to ensure this
equality. In addition, for gap metrics, one can also quantify the degree of
suppression of the $T$ parameter, \ \
\begin{eqnarray}
  \alpha T & = & \mathcal{O} \left( G^{- 3} \right) . 
\end{eqnarray}
\subsection{The Fermi constant}\label{fermi}

The expression of the Fermi constant in this class of models is given by
\begin{eqnarray}
  G_F & \equiv & \sqrt{2}  \sum_{X = V, A ; n = W}^{\infty}
  \frac{A_{X^{\pm}_n} \left( l_0 \right)^2}{\left( M_{X^{\pm}_n} \right)^2}, 
  \label{GF2}
\end{eqnarray}
with the sum running over the $W$ and its KKs, $n = W, 1, \ldots,
\infty$ and where $A_{X^{\pm}_n}$ is the axial component of $\left| \Psi_{X^{\pm}_n}\right\rangle $.
This is an example of the sums considered in (\ref{gnM2}), this time with UV
BCs that mix $V$ \& $A$ fields. One must again express $A \left( z \right) /
M^2$ as a double integral by integrating twice the EOM,
\begin{eqnarray}
  \frac{A_{X_n^{\pm}} \left( z \right)}{M_{X_n^{\pm}}^2} & = & \int_z
  \frac{1}{w_A}  \left( \int^{z'} w_A A_{X_n^{\pm}} - \int w_V V_{X_n^{\pm}}
  \right) .  \label{AoM2}
\end{eqnarray}
With this, we obtain
\begin{eqnarray}
  G_F & = & \sqrt{2}  \sum_{X = V, A ; n = W}^{\infty} A_{X_n^{\pm}} \left(
  l_0 \right)  \int \frac{1}{w_A}  \int^z w_A A_{X_n^{\pm}}, 
\end{eqnarray}
where the second term in (\ref{AoM2}) vanishes when inserted into (\ref{GF2}).
Finally, using the completeness relation we get
\begin{eqnarray}
  G_F & = & \frac{1}{\sqrt{2} g_5^2}  \int \frac{1}{w_A} . \bignone 
\end{eqnarray}
From Equation (\ref{fpi}), we can rewrite the result in terms of the decay
constant of the would-be Goldstones that are eaten
\begin{eqnarray}
  G_F & = & \frac{1}{\sqrt{2} f^2} .  \label{GF}
\end{eqnarray}
This is an exact relation obtained through SRs and shows that the would-be
Goldstones are not eaten solely by the $W$ mode, but are also fed partly to
the KKs. This result is very intuitive but, from the 4D perspective, Equation
(\ref{GF}) could be verified only after taking into account {\tmem{all }}KK
contributions. \

\subsection{The rho parameter}\label{rho-param}

The low-energy rho parameter in this class of models reads
\begin{eqnarray}
  \rho_{\ast} \left( 0 \right) & \equiv & \left. \left. \sqrt{2}  \sum_{X = V,
  A, B ; n = Z}^{\infty} \frac{A_{X_n^0} \left( l_0 \right)^2}{M^2_{X_n^0}}
  \right/ G_F \right.,  \label{rho}
\end{eqnarray}
where the sum includes the $Z$ and all neutral KK modes. Due to the custodial
symmetry, the expression for $A / M^2$ (\ref{AoM2}) is still valid for the
neutral fields {\footnote{Note that this is only true for the axial
component.}}. From there on, the derivation goes through from the charged to
the neutral case, mutatis mutandi {\footnote{The massless photon can be
omitted in the application of the completeness relation, since it has no axial
component.}}
\begin{eqnarray}
  \sum_{X = V, A, B ; n = Z}^{\infty} \frac{A_{X_n^0} \left( l_0
  \right)^2}{M^2_{X_n^0}} & = & \sum_{X = V, A, B ; n = Z}^{\infty} A_{X_n^0}
  \left( l_0 \right)  \int \frac{1}{w_A}  \int^z w_A A_{X_n^0} \\
  & = & \frac{1}{2 f^2}, 
\end{eqnarray}
and thus
\begin{eqnarray}
  \rho_{\ast} \left( 0 \right) & = & 1,  \label{rho1}
\end{eqnarray}
{\tmem{independently of the geometry}} $w_V (z), w_A \left( z \right)$. The
only requirement to obtain $\rho_{\ast} \left( 0 \right) = 1$ is the bulk
$\tmop{SU} (2) \times \tmop{SU} (2)$ symmetry. This is again a good example of
how 5D symmetries show up in 4D observables by (not) summing over the whole
tower of KKs.

For the case of bulk fermions, the expression (\ref{rho}) would then involve
the square of the overlap integrals of the fermions with $A$. The same
modification occurs for the $W$ tower as for the $Z$ tower, leading again to
(\ref{rho1}).

\subsection{The $T$ parameter}

The $T$ parameter, defined through
\begin{eqnarray}
  1 + \alpha T & = & \left. \frac{A_Z \left( l_0 \right)^2}{M_Z^2} \right/ \frac{A_W \left(
  l_0 \right)^2}{M_W^2}, 
\end{eqnarray}
only involves two eigenmodes, the $W$ and $Z$. $T$ can be calculated by
subtracting the contributions of KK modes from the total sums of Section
\ref{fermi}, writing

\begin{eqnarray}
  \frac{A_W \left( l_0 \right)^2}{M_W^2} & = & \frac{1}{2 f^2} - \sum_{X = V,
  A ; n = 1}^{\infty} \frac{A_{X^{\pm}_n} \left( l_0
  \right)^2}{M_{X^{\pm}_n}^2},  \label{AWM}\\
  \frac{A_Z \left( l_0 \right)^2}{M_Z^2} & = & \frac{1}{2 f^2} - \sum_{X = V,
  A, B ; n = 1}^{\infty} \frac{A_{X^0_n} \left( l_0 \right)^2}{M_{X^0_n}^2} . 
  \label{AZM}
\end{eqnarray}
In gap metrics, the KK contributions are suppressed with respect to those of
the UL because
\begin{itemize}
  \item the KKs masses are enhanced with respect to that of $W$ and $Z$
  \begin{eqnarray}
    \frac{M_{W, Z}^2}{M_{\tmop{KK}}^2} & = & \mathcal{O} \left( G^{- 1}
    \right), 
  \end{eqnarray}
  \item the KK wave-functions are repelled from the brane, whereas those of
  the UL modes are not
  \begin{eqnarray}
    \frac{A_{\tmop{KK}} \left( l_0 \right)^2}{A_{W, Z} \left( l_0 \right)^2} &
    = & \mathcal{O} \left( G^{- 1} \right),
  \end{eqnarray}
  \item the KK masses and wave-functions are approximately isospin symmetric
  \begin{eqnarray}
    M_{X^0_n}^2 / M_{X^{\pm}_n}^2 & = & 1 +\mathcal{O} \left( G^{- 1} \right),
    \\
    A^2_{X_n^0} \left( l_0 \right) / A^2_{X_n^{\pm}} \left( l_0 \right) & = &
    1 +\mathcal{O} \left( G^{- 1} \right), \\
    A^2_{B_n} \left( l_0 \right) & = & 0 +\mathcal{O} \left( G^{- 3} \right), 
  \end{eqnarray}
\end{itemize}
where $X = V, A$. The last two results can be established using equation
(\ref{AoM2}). Since each factor introduces a suppression by $1 / G$, we get
\begin{eqnarray}
  \alpha T & = & \mathcal{O} \left( G^{- 3} \right) . 
\end{eqnarray}

\section{Conclusions}

What distinguishes a 5D model from a 4D model of resonances? An effective
description of a 5D model inherits interesting properties from its
extra-dimensional nature. In other words, {\tmem{all}} the KKs
{\tmem{cooperate}} together to preserve 5D symmetries and a soft behavior at
high energies.

To see this characteristic behavior one needs a method to shortcut the
{\emdash}impossible{\emdash} task of summing over an infinite number of KKs.
In this paper we have presented such a shortcut that follows the ideas of
holography.

One advantage of this method is that infinite sums over KKs are converted into
{\tmem{geometrical factors}}: integrals involving just the metric, and not the
individual KK properties. These geometrical factors lead to analytic
expressions for physical observables.

Using this method, properties that depend on the 5D nature of the model
(non-locality, 5D symmetries) are automatically verified,
{\tmem{independently}} of the particular metric considered in the model.
Another use of the method is to classify 5D models (metrics) in terms of their
approximate behavior. In this paper, we identify a subclass of metrics,
{\tmem{gap metrics}}, suitable to be a good description for electroweak
symmetry breaking.

Finally, the most interesting physical quantities are quadratic. Even in the
presence of background fields, our method is valid to obtain quadratic
observables like the $S, T$ or $\rho_{\ast}$. This is possible thanks to the rewriting of
these background fields as effective metrics {\cite{hep-ph/0612239}}.

\begin{acknowledgments}
J.H. is supported by DOE grant
DE-FG02-92ER40704 and V.S. by grant DE-FG02-91ER40676.

\end{acknowledgments}
\appendix

\section{Notations}\label{notations}

We will use the following shorthand notation
\begin{eqnarray}
  \int_{l_0}^{l_1} \mathd zf (z) \bignone  & \equiv & \int f,\\
  \int_{l_0}^z \mathd z' f (z') \bignone  & \equiv & \int^z f \bignone,\\
  \int_z^{l_1} \mathd z' f (z') \bignone & \equiv & \int_z \bignone f,
\end{eqnarray}
where we omit the integration variables and the limits if they lie on the
interval endpoints $(l_0, l_1)$. \ Double integrals are denoted as follows,
\begin{eqnarray}
  \int_{l_0}^z f (z') d z' \bignone  \int_{l_0}^{z'} g (z'') d z'' & = &
  \int^z f \int^{z'} g .
\end{eqnarray}
\section{Mixed IR BC}\label{App-mixed}

In most of the paper, we present results for $(+)$ or $\left( - \right)$ IR
BCs. The case of mixed BC can also be treated in the same manner, and we give
an example here. This is particularly relevant since the effect of a coupling
to a bulk scalar obtaining a vev (\ref{bgfield}) also effectively yields a
mixed BC, as shown in {\cite{hep-ph/0612239}}.

As an example, we compute the sum $\sum 1 / M_n^2 \bignone$ for the case with
mixed IR BC (and $\left( - \right)$ UV BC). We first point out that the EOM
can always be rewritten as
\begin{eqnarray}
  - \frac{1}{w \alpha_{\kappa}^2} \partial \left( w \alpha_{\kappa}^2 \partial
  \frac{\varphi_{\kappa}}{\alpha_{\kappa}} \right) & = & M^2 
  \frac{\varphi_{\kappa}}{\alpha_{\kappa}},  \label{modifEOM}
\end{eqnarray}
where $w$ is the effective metric for the modes in question, and
$\varphi_{\kappa}$ satisfies the IR BC
\begin{eqnarray}
  \left. \partial \left( \log \varphi_{\kappa} \right) \right|_{l_1} & = & -
  \kappa . 
\end{eqnarray}
$\kappa = 0, \infty$ corresponds to $(+)$ and $(-)$ IR BCs. Anything in between
corresponds to mixed BCs.

To make sense of (\ref{modifEOM}), we have to recall the definition of the
auxiliary function $\alpha$. It is defined to satisfy a massless equation
\begin{eqnarray}
  \partial \left( w \partial \alpha_{\kappa} \right) & = & 0, 
\end{eqnarray}
and the same IR BC as $\varphi_{\kappa}$, i.e.
\begin{eqnarray}
  \left. \partial \left( \log \alpha_{\kappa} \right) \right|_{l_1} & = & -
  \kappa , 
\end{eqnarray}
but normalized to one on the UV brane
\begin{eqnarray}
  \left. \alpha_{\kappa} \right|_{l_0} & = & 1 . 
\end{eqnarray}
As shown in {\cite{hep-ph/0612239}}, the function $\alpha_{\kappa}$ can be
determined explicitly as
\begin{eqnarray}
  \alpha_{\kappa} & = & \frac{\frac{1}{\kappa w \left( l_1 \right)} + \int_z
  \frac{1}{w} \bignone}{\frac{1}{\kappa w \left( l_1 \right)} + \int
  \frac{1}{w}} .  \label{alphageneral}
\end{eqnarray}

There are then two cases: $(+)$ and mixed IR BCs on one side, which can be
rewritten as
\begin{eqnarray}
  \left. \partial \left( \frac{\varphi}{\alpha} \right) \right|_{l_1} & = & 0,
  \label{phialpBC}
\end{eqnarray}
and $(-)$ IR BC, for which
\begin{eqnarray}
  \left. \alpha \right|_{l_1} & = & \left. \varphi \right|_{l_1} \hspace{1em}
  = \hspace{1em} 0 . 
\end{eqnarray}
For the first case (\ref{phialpBC}), one can directly apply the technique of
Section \ref{what} on the function $\varphi / \alpha$, since it obeys a
standard EOM with effective metric $w \alpha^2$, and with $\left( -, + \right)$ BCs.
This yields
\begin{eqnarray}
  \text{for } \kappa < \infty, \hspace{2em} \frac{\varphi_{\kappa,
  n}}{M_{\kappa, n}^2} & = & \alpha_{\kappa}  \int^z \frac{1}{w
  \alpha_{\kappa}^2}  \int_{z'} w \alpha_{\kappa} \varphi_{\kappa, n} . 
\end{eqnarray}
This implies
\begin{eqnarray}
  \sum_{n = 1}^{\infty} \frac{1}{M_{\kappa, n}^2} & = & \int w
  \alpha_{\kappa}^2  \int^z \frac{1}{w \alpha_{\kappa}^2} . \bignone 
\end{eqnarray}
Note that this reproduces (\ref{resSR1}) for $\kappa = 0$, in which case
$\alpha_0 = 1$. For the $\left( -, - \right)$ case, one pays attention to the
boundary terms when integrating the EOM twice. This yields
\begin{eqnarray}
  \frac{\varphi_{-, n}}{M_{-, n}^2} & = & \left( \alpha_- - 1 \right)  \int
  \frac{1}{w}  \int_{z'} w \varphi_{-, n} + \int^z \frac{1}{w}  \int_{z'} w
  \varphi_{-, n}, 
\end{eqnarray}
and thus
\begin{eqnarray}
  \sum_{n = 1}^{\infty} \frac{1}{M_{-, n}^2} & = & \int w \alpha_-  \int^z
  \frac{1}{w}, 
\end{eqnarray}
as promised in (\ref{resSR1p}).

\bibliography{new-sum-rules}
\bibliographystyle{apsrev}

\end{document}